\documentclass{CHEP2004}
\usepackage{graphicx}

\newcommand{\ifb}{fb\ensuremath{{}^{-1}}}

\begin{document}
\title{New distributed offline processing scheme at Belle}

\author{F. J. Ronga, I. Adachi, N. Katayama, KEK, Tsukuba, Japan}

\maketitle

\begin{abstract}
The offline processing of the data collected by the Belle detector has
been recently upgraded to cope with the excellent performance of the
KEKB accelerator. The 127~fb${}^{-1}$ of data (120~TB on tape)
collected between autumn 2003 and summer 2004 has been processed in 2
months, thanks to the high speed and stability of the new, distributed
processing scheme. We present here this new processing scheme and its
performance.
\end{abstract}

\section{INTRODUCTION}
The Belle experiment~\cite{belle}, located on the KEKB~\cite{kekb}
asymmetric-energy $e^+e^-$ collider, is primarily devoted to the study
of CP violation in the $B$~meson system.  KEKB has shown a very stable
operation with increasing luminosity over the years. It has turned to
so-called ``continuous injection mode'' last January, thus allowing a gain
in the integrated luminosity of about 30\%. In this mode, the beam particle
losses are compensated by continuously injecting beam from the linear
accelerator, without interruption of data taking.

KEKB has reached a world record peak luminosity of 
$1.39\times10^{34}$~cm${}^{-2}$s${}^{-1}$ and an integrated luminosity
of about 1~\ifb\ per day. In the meanwhile, Belle has
accumulated a total integrated luminosity of 290~\ifb\ (about
275~million $B$~meson pairs), among which 127~\ifb\ was collected
between October~2003 and July~2004 (SVD2 run).

\begin{figure}[htb]
\centering
\includegraphics*[width=0.45\columnwidth]{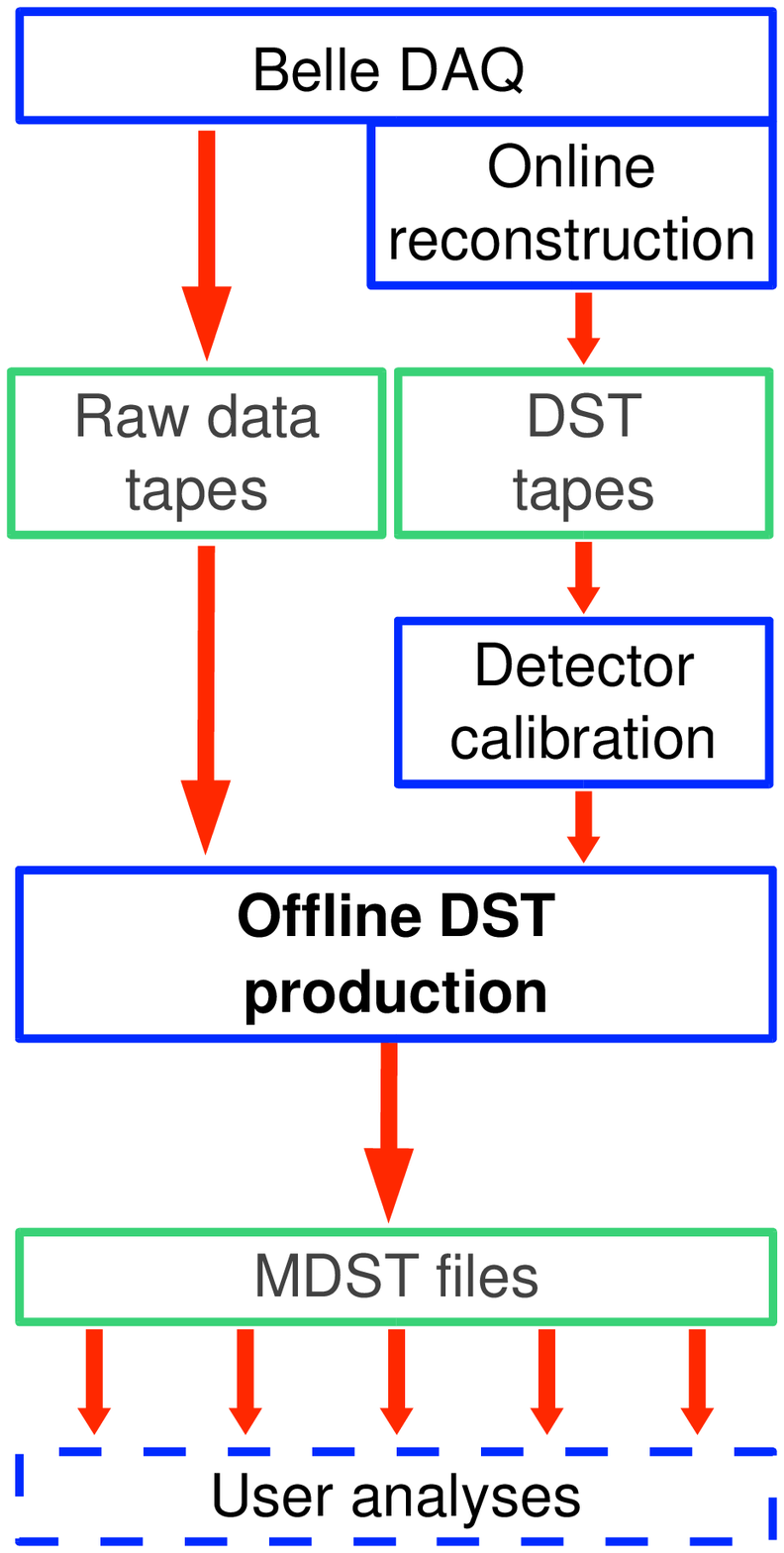}
\caption{The Belle data flow.}
\label{flow-fig}
\end{figure}

\begin{table}[hbt]
\begin{center}
\caption{Belle data acquisition and processing figures. 
Numbers in parentheses correspond to data accumulated during 
SVD2 run.}
\begin{tabular}{|lll|}
\hline
DAQ output rate    & 8.9~MB/s &\\
\hline
Raw event size & 38~kB &\\
DST event size & 60~kB &\\
mdst event size & 12~kB& \\
\hline
Total raw data & 247~TB &(120~TB)\\
Total DST data & 390~TB &(186~TB)\\
Total mdst data & 80~TB &(40~TB)\\
\hline
\end{tabular}
\label{flow-tab}
\end{center}
\end{table}

A (simplified) overview of the Belle data flow is shown in 
Figure~\ref{flow-fig}. The subdetector information is collected by the Belle 
data acquisition (DAQ) at an average output rate of about 230 events per
second. This raw data is stored on tapes and is, at 
the same time, processed by the online reconstruction farm~\cite{rfarm} (RFARM). 
The data processed by the RFARM is stored to data summary
tapes (DST) and is used for detector calibration. The offline DST
processing then reads in the raw data tapes and uses the calibration constants
to produce mini-DST (mdst) files, which are used for physics analyses.
The main figures of the data flow are summarized in Table~\ref{flow-tab}.

In this paper, we present the new offline processing scheme used to process
the data collected during SVD2 run. We first introduce the computing hardware
and software tools. An overview and the performance of the processing scheme
are then given.

\section{COMPUTING HARDWARE}

The hardware used for DST processing consists of storage systems
and computing farms detailed below. The components are connected
together by Gigabit ethernet switches.

\subsection{Data Storage}

Two types of data storage are used.

Raw and DST data is stored on
SONY DTF2 tapes (200~GB tapes), providing a total storage space of
500~TB. They are accessed through 20 tape servers (Solaris mainframe
servers with 4 CPUs at 0.5~GHz each), each server being connected to 
two tape drives with a maximum readout rate of 24~MB/s. 

Mdst data is stored on a new hierarchical storage management 
system~\cite{HSM} (HSM) provided by SONY. It consists of a hybrid 
disk and tape storage
system with 500~GB tapes and 1.6~TB RAID disks. The total tape storage
space is 450~TB (to be soon expanded to 1.2~PB), while disk space reaches
26~TB in total. The tapes are readout by SONY SAIT tapes with a maximum
rate of 30~MB/s. The 16 disks are connected to 8 servers with 4~CPUs
at 2.8~GHz each. The data stored on disk is automatically migrated to 
the tapes by the HSM system. Unused data is deleted from the disks
and automatically reloaded when accessed by users.

\subsection{Computing Farms}

The DST processing was performed on three classes of PC farms:
\begin{Itemize}
\item Class I farm: 60 hosts with 4 Intel Xeon CPUs at 0.766~GHz each 
      (16.4\% of total power).
\smallskip

\item Class II farm: 119 hosts with 2 Intel Pentium III CPUs at 1.26~GHz each
      (26.7\% of total power).
\smallskip

\item Class III farm: 100 hosts with 2 Intel Xeon CPUs at 3.2~GHz each
      (56.9\% of total power).
\end{Itemize}

The total available CPU power is 1.12~THz, distributed among 279~hosts.
These are divided into 8 different clusters of 30 to 40 hosts.

\section{SOFTWARE TOOLS}

The DST processing scheme makes use of a number of software tools.
The core software is based on three ``home-grown'' tools detailed
below. Other tools are also mentioned.

\begin{figure}[htb]
\centering
\includegraphics*[width=0.6\columnwidth]{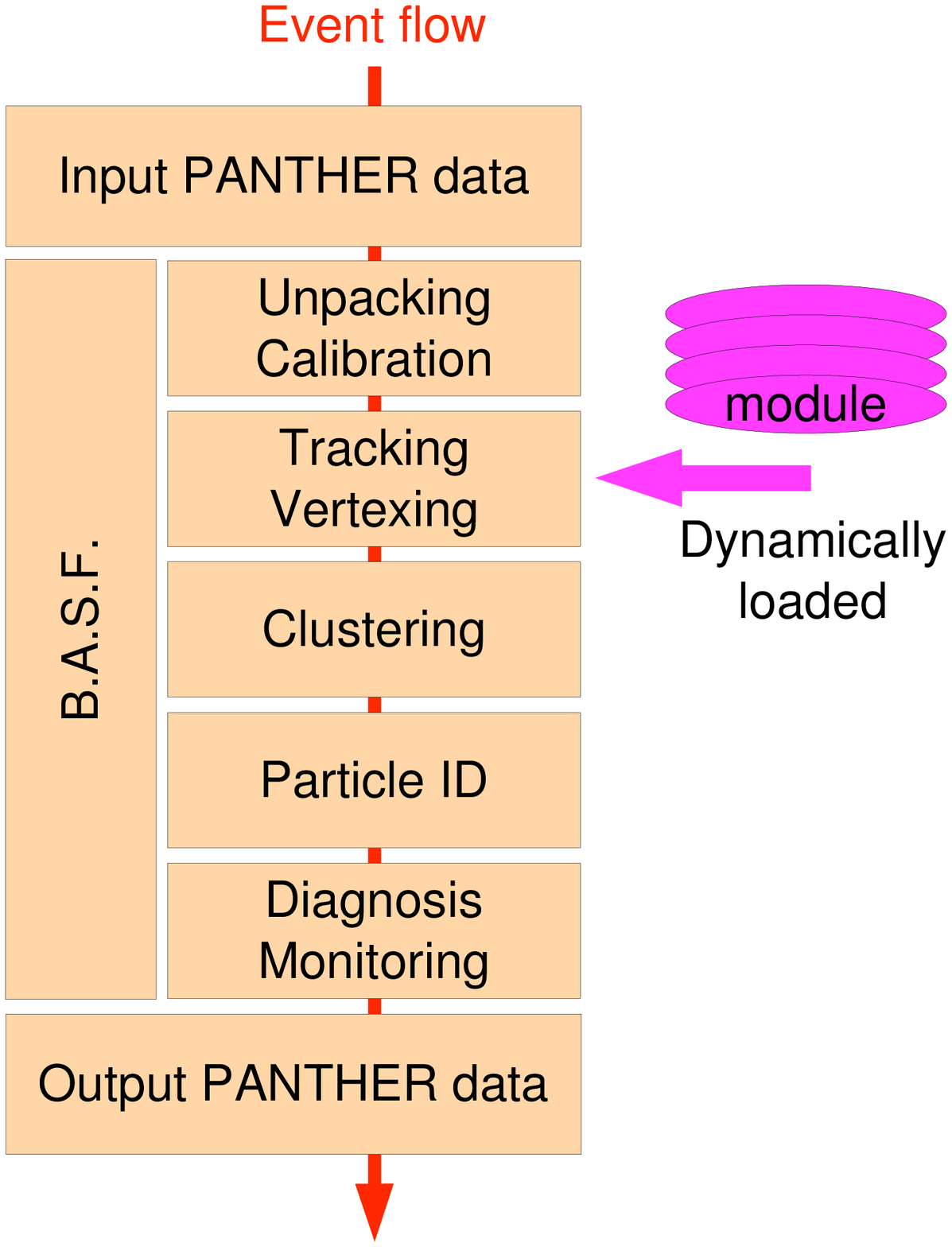}
\caption{Schematic view of the event flow in DST processing.}
\label{basf-fig}
\end{figure}

\subsection{The Belle Analysis Software Framework}

B.A.S.F. is the software framework used for all Belle data analyses, 
from data acquisition to end-user analysis and DST processing.
It provides an interface to external programs (modules), 
dynamically loaded as shared objects at the start of a processing job. 
The interface includes begin and end run calls, event calls, 
histogram definitions, as well as a shared memory utility.
External modules actually process the event information. 
Several modules can be called at will, in the order specified by the user.
B.A.S.F. is written in C++ (and so are the modules).
Finally, B.A.S.F. supports Symmetrical Multiprocessing (SMP), 
thus allowing parallel processing of events on a multi-processor machine.
Figure~\ref{basf-fig} shows the event flow for a DST processing job.

\subsection{Network Shared Memory}

The NSM package provides tools for information exchange over a 
TCP/IP based LAN. It allows processes running on different machines 
to share memory across the network, or send requests and messages 
to each other.

\begin{figure}[hbt]
\centering
\includegraphics*[height=\columnwidth,angle=90]{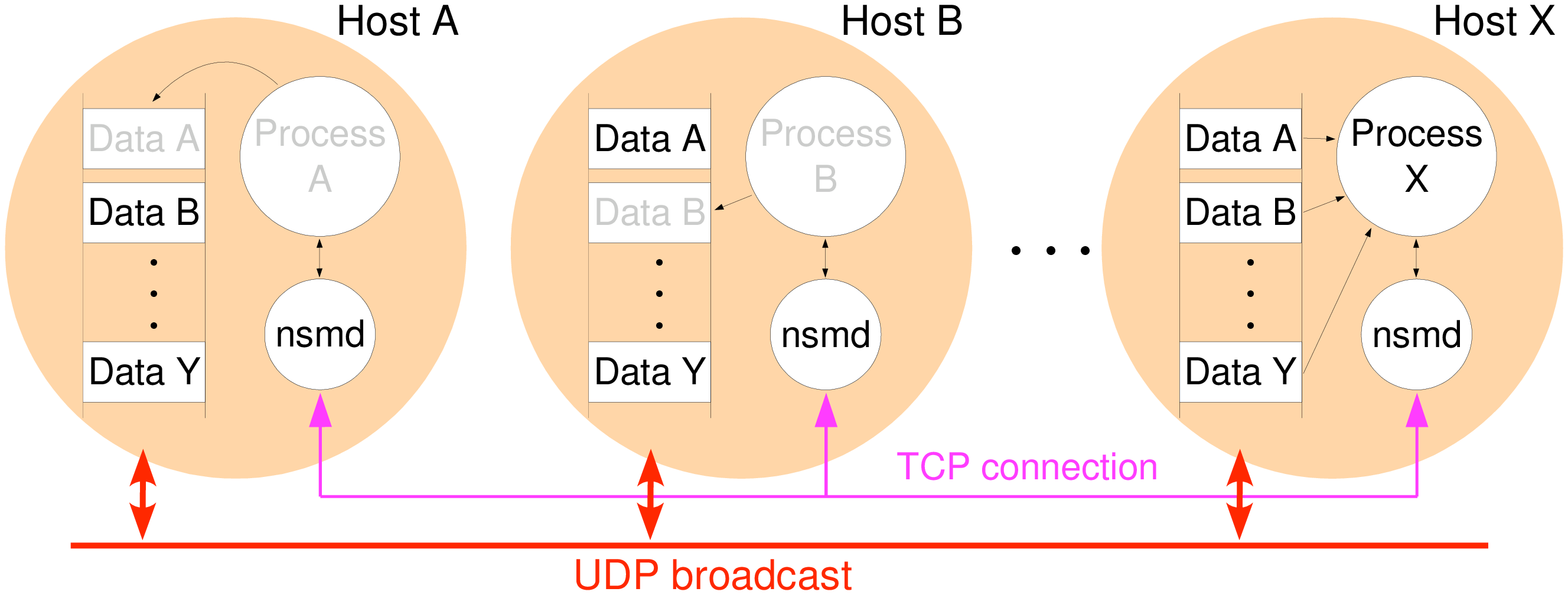}
\caption{The Network Shared Memory.}
\label{nsm-fig}
\end{figure}

\subsection{The PANTHER format}

The input and output of Belle data, as well as transfer between modules, 
is managed by the PANTHER system. It is used consistently from raw data 
to user analysis. The PANTHER format consists of compressed tables (banks), 
using the standard zlib libraries. A cross-reference system is implemented 
in order to allow navigation between the tables. The table formats are 
defined in B.A.S.F. in ASCII header files, which are loaded before the 
modules. Users may define their own tables.

\subsection{Other tools}

In addition to the software described above, DST processing uses various
other standard tools:

\begin{Itemize}
\item The postgresql database system~\cite{psql}. An important part
of the information relevant to DST processing is stored in databases:
calibration constants; meta-data information about raw data, DST and 
mdst files; tapes and tape drives information; PC farms description.
The DST processing uses a dedicated postgresql server mirrored from the
main database server.
\smallskip

\item LSF batch queues~\cite{LSF}. Tape servers are operated through the LSF
queuing system.
\smallskip

\item Redhat~\cite{redhat} and Solaris~\cite{solaris}. The computing farms
run on various versions of Redhat Linux: Redhat~6.2 (class I), 7.2 (class II)
or 9.0 (class III). The tape servers run Solaris (SunOS 5.7).

\end{Itemize}

\section{DST PROCESSING SCHEME}

The new DST processing scheme has been implemented between 
March and May 2004. It is based on a distributed version of B.A.S.F.,
dbasf, first developed for the online processing (RFARM) and adapted 
to the configuration of offline processing. The computing facility used 
for DST processing was divided in a number of dbasf clusters (see
sub-section {\it Computing Farms} above) that 
independently process groups of events (runs). The processing is fully 
managed by a steering Perl script (dcruncher) that allocates jobs and 
surveys the global DST processing operations.

\subsection{Distributed B.A.S.F.}

In order to increase its parallel-processing ability, B.A.S.F. was 
extended to a distributed version that relies on the NSM system. 
A dbasf cluster physically consists of a tape server and 30 to 40 
PC hosts (PC cluster), as shown on Figure~\ref{dbasf-fig}.
The data is distributed by an input node (running on a tape server) 
to all PC hosts (basf nodes), on which basf processes are running. 
The output is redirected to a single output node (one of the PC hosts) 
that sends it to the HSM storage system. Finally, the synchronization 
among the various nodes is managed by a master node running on 
one of the PC hosts. 

\begin{figure}[htb]
\centering
\includegraphics*[height=0.8\columnwidth,angle=90]{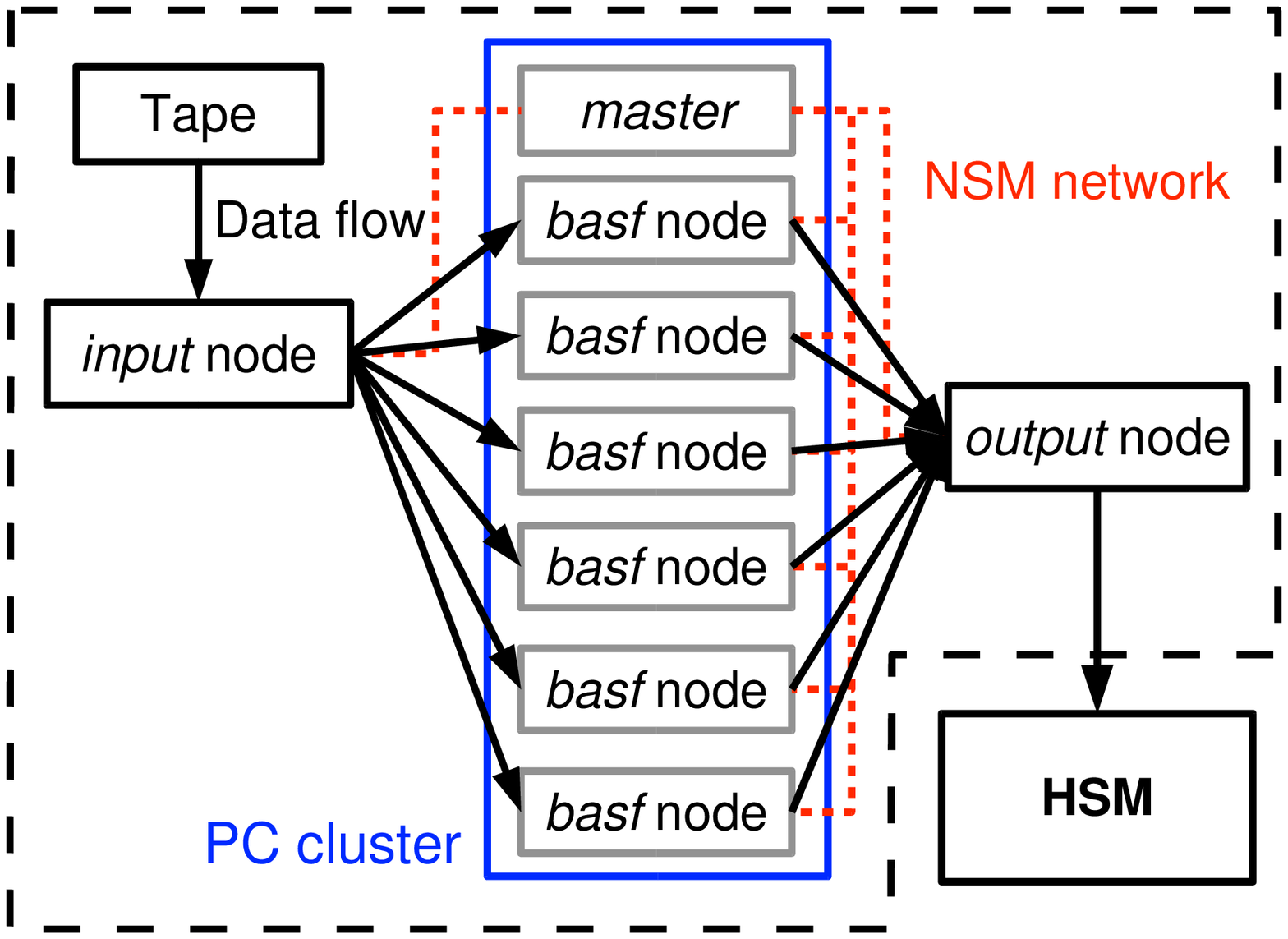}
\caption{A distributed B.A.S.F. cluster.}
\label{dbasf-fig}
\end{figure}

\subsection{Processing Stream}

The processing is driven by the {\it dcruncher} Perl script that runs on a 
mainframe Solaris server (similar to tape servers). In order to allocate 
processing jobs, it interacts with the database dedicated to DST processing.
A script is sent to the LSF batch queue to run the job.
During the job, the LSF script checks the operation of the dbasf cluster. 
After completion of a job, dcruncher checks that output files exist and have 
sensible sizes. In case of any failure, the DST team is immediately 
informed by e-mail.

\begin{figure}[htb]
\centering
\includegraphics*[height=0.8\columnwidth,angle=90]{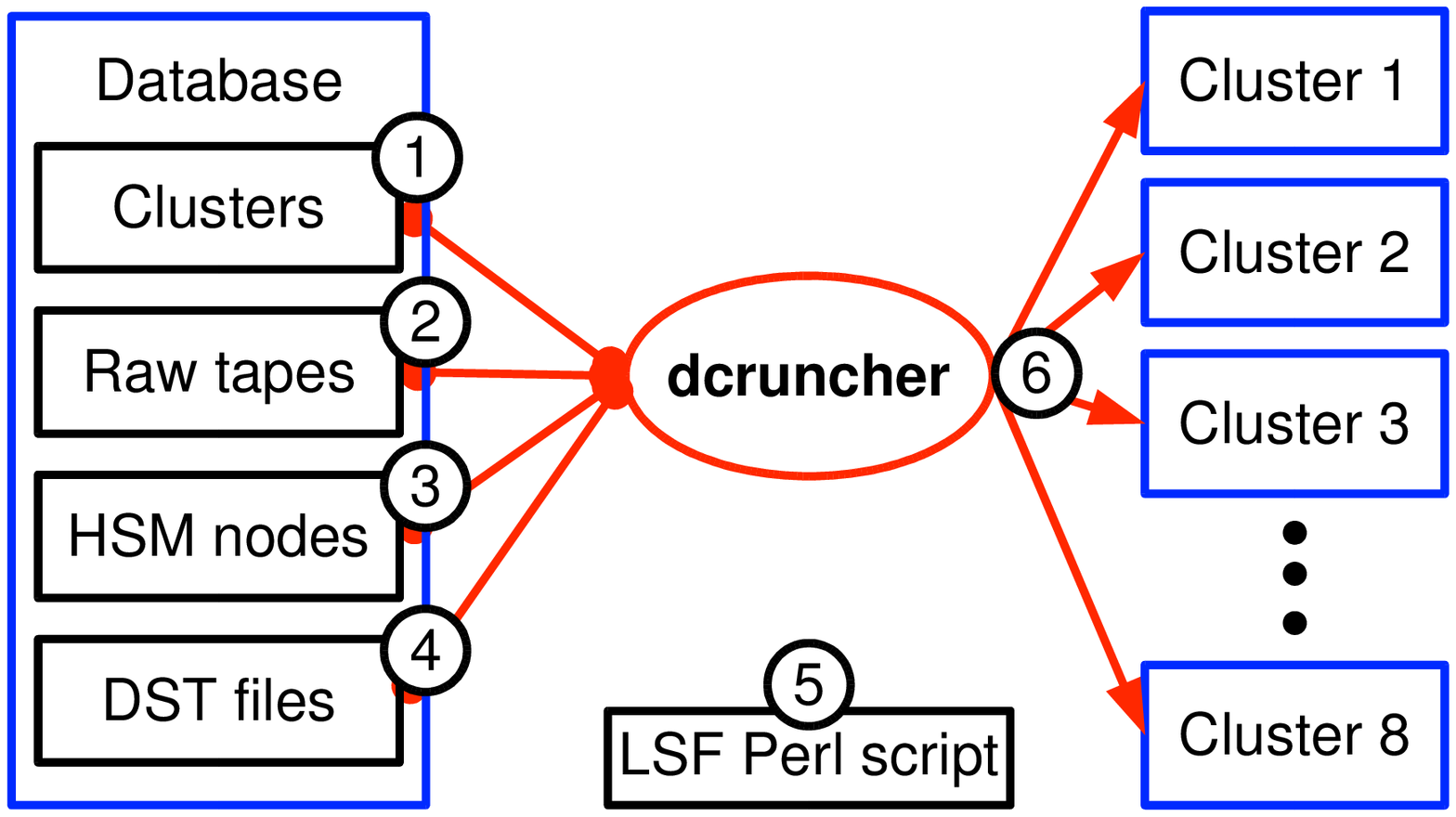}
\caption{Block diagram showing the processing stream.}
\label{dcruncher-fig}
\end{figure}

The dcruncher script runs through the following steps (see 
Figure~\ref{dcruncher-fig}):
\begin{enumerate}
\item Check for free dbasf clusters (fastest first).

\item Check for available unprocessed raw tape (depending 
      on cluster type: see speed optimization below).
      
\item Find location for output file on HSM.

\item Insert corresponding entry in mdst files table.

\item Write Perl script job to submit to LSF batch queue.

\item Send job to relevant cluster (the script job is run 
      on the tape server of the cluster).
      
\item Wait 5 minutes and restart the loop.

\end{enumerate}

\subsection{Speed Optimization}

The number of events in each run greatly varies,
depending on the operation condition. In order to optimize the
allocation of computing power, faster clusters process the largest 
runs, and slower clusters process the smallest runs. 
The total process time is indeed roughly proportional
to the run size. This simple algorithm is illustrated on Figure~\ref{events-fig}: 
the largest runs are processed by class~III clusters,
while shortest runs are processed by class~I clusters. 
The number of events in runs processed by class~II clusters 
is randomly distributed between these two extrema, because these 
clusters simply process runs sequentially.

\begin{figure}[hbt]
\centering
\includegraphics*[width=\columnwidth]{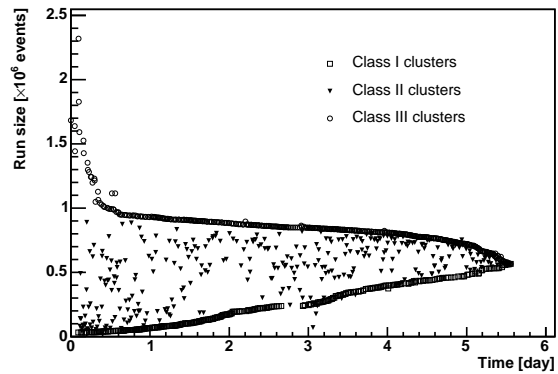}
\caption{Time evolution of the number of events per run during the
processing (see text for details).}
\label{events-fig}
\end{figure}

\section{PERFORMANCE}

\begin{figure}[hbt]
\centering
\includegraphics*[width=\columnwidth]{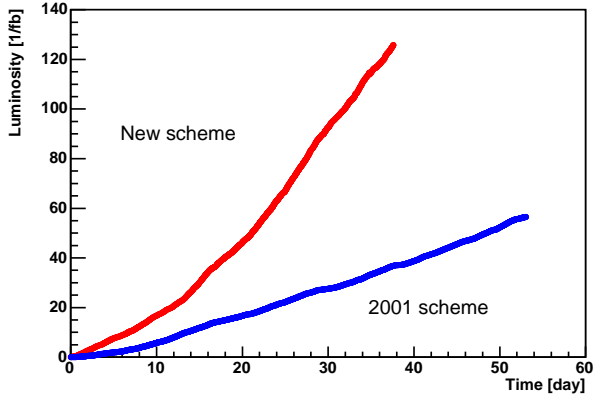}
\caption{Performance comparison between the old and the new scheme. 
The dead-time periods due to delays independent of the DST processing 
are subtracted.}
\label{DSTsum-fig}
\end{figure}

The task of the new scheme was to complete the processing of all
the data accumulated from autumn 2003 to summer 2004 (120~TB) 
before the summer physics conferences. It started running on May 18,
2004 and finished processing on July 12 (56 days in total). It has
been actually running for 39 days (delays were mainly due to calibration
and updating of the database). 3.3~billion events were processed in
3408 jobs, with a total output size of 40~TB. The average output rate
was 3.2/fb/day (37~MB/s), with a peak of 5.1/fb/day (60~MB/s) for
7~consecutive days. In comparison, the previous processing scheme
reached an average output rate of 1.1/fb/day~(see Figure~\ref{DSTsum-fig}).
This older scheme used a slightly different version of dbasf on about
half of the computing power. The output rate per cluster class is shown
in Figure~\ref{speed-fig}.

\begin{figure}[htb]
\centering
\includegraphics*[width=\columnwidth]{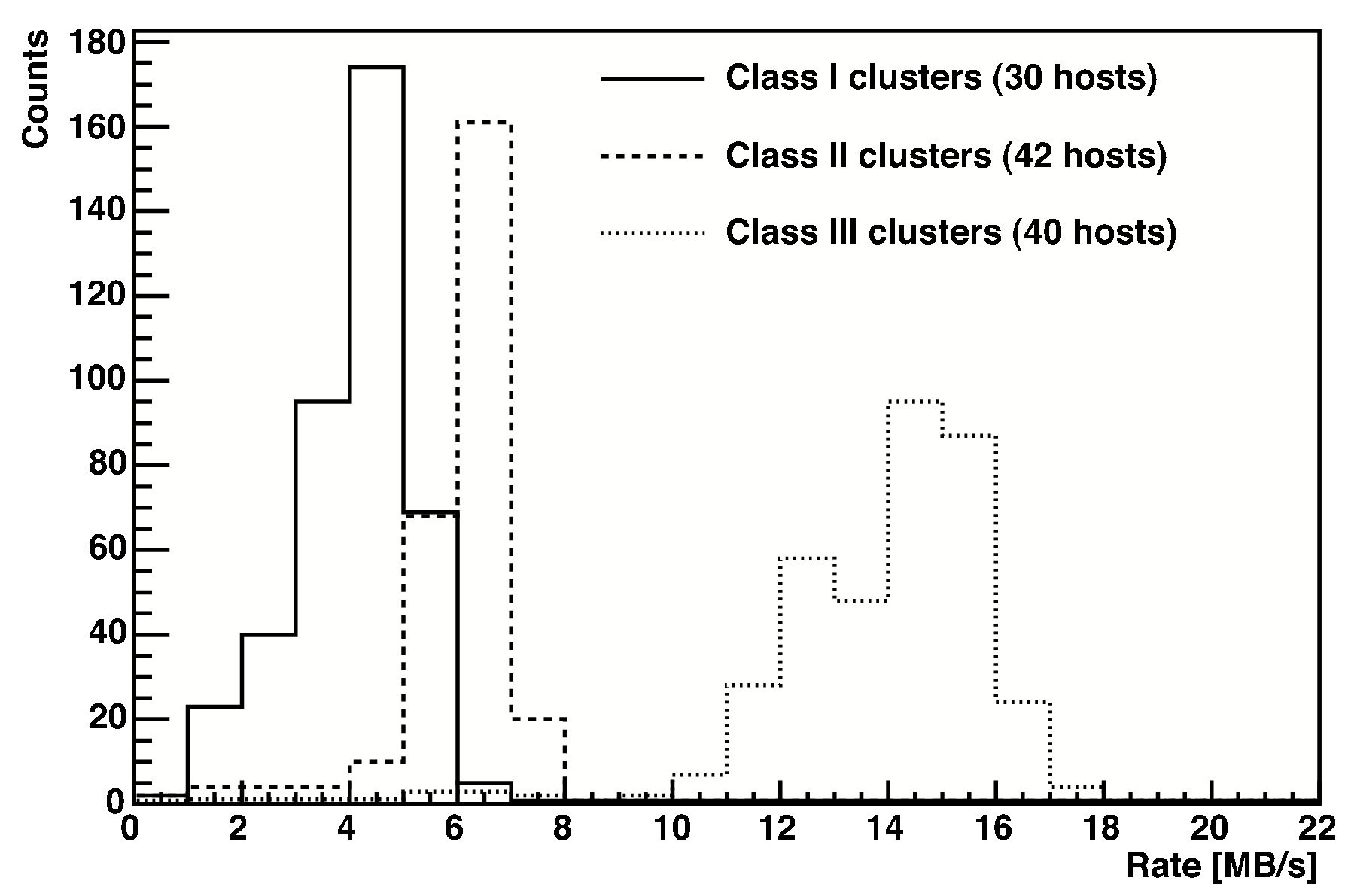}
\caption{Processing rate for the various cluster classes.}
\label{speed-fig}
\end{figure}

\subsection{Failures and limitations}

During the 39 days of processing, a total of 0.7\% of the jobs
failed due to an error related to dbasf. The break-down of the
errors is shown in Table~\ref{failure-tab}. 

\begin{table}[hbt]
\begin{center}
\caption{Frequency of the various possible failure causes.}
\begin{tabular}{|lr|}
\hline
Inter-process communication & 0.2\%    \\
Database access             & 0.1\%    \\
Tape drives                 & $<0.1\%$ \\
Network                     & 0.3\%    \\
\hline
Total                       & 0.7\%    \\
\hline
\end{tabular}
\label{failure-tab}
\end{center}
\end{table}

In addition, a number of limitations to the processing speed
have been observed:

\begin{Itemize}
\item Database access is one of the main issues, since processing makes 
      heavy use of the database, in particular at the start of a job.
      Using more dedicated servers will solve this issue.

\item The limited CPU power of tape servers which distribute the data 
      the dbasf clusters was identified as the present bottleneck.
      Faster tape servers will be used in the future.

\item The network bandwidth between the input server and basf nodes may 
      eventually limit the processing power. This does not seem to be
      a major issue in the near future.

\end{Itemize}

None of these limitations, however, seriously hampered
the processing speed.

\section{CONCLUSION}

The new offline processing scheme started running on May 18, 2004. In 39 
days of stable running, it successfully processed the 3 billion events 
collected by KEKB with a maximum rate of 5/fb/day, 5 times faster than 
the data acquisition and the previous scheme. With the expected increase 
of the KEKB luminosity, however, the DST processing will face further 
challenges\dots Room for improvement of this scheme still remains.


\begin{thebibliography}{9}   

\bibitem{belle} A.~Abashian et al., Nucl. Instr. and Meth. A479, 117 (2002).

\bibitem{kekb} S.~Kurokawa et al., Nucl. Instr. and Meth. A499, 1 (2003).

\bibitem{rfarm} R.~Itoh, ``Experience with Real Time Event Reconstruction Farm for Belle Experiment'', Contribution 209, CHEP'04.

\bibitem{HSM} N.~Katayama, ``New compact hierarchical mass storage system at 
Belle'', Contribution 211, CHEP'04.

\bibitem{psql} http://www.postgresql.org/

\bibitem{LSF} Platform Computing -- http://www.platform.com/

\bibitem{redhat} http://www.redhat.com/

\bibitem{solaris} http://www.sun.com/software/solaris/

\end{thebibliography}
\end{document}